\documentclass{kapproc} 

\usepackage[dvips]{graphicx}

\newcommand{\one}{$\rm (1-0)$}
\newcommand{\two}{$\rm (2-1)$}

\newcommand{\bdm}{\begin{displaymath}}
\newcommand{\edm}{\end{displaymath}}
\newcommand{\beq}{\begin{equation}}
\newcommand{\eeq}{\end{equation}}
\newcommand{\eit}{\end{itemize}}
\newcommand{\ben}{\begin{enumerate}}
\newcommand{\een}{\end{enumerate}}
\newcommand{\bfi}{\begin{figure}[htb]}
\newcommand{\bpfi}{\begin{figure}[p]}

\upperandlowercase

\kluwerbib 

\begin{document}



\articletitle[]{The far-infrared properties of the most isolated galaxies}


\author{U. Lisenfeld\altaffilmark{1,2}, L. Verdes-Montenegro\altaffilmark{2}, 
S. Leon\altaffilmark{2}, and J. Sulentic\altaffilmark{3}}

\altaffiltext{1}{Dept. F\'{i}sica Te\'{o}rica y del Cosmos, Universidad de Granada, Spain}
\altaffiltext{2}{Instituto de Astrof\'\i sica de Andaluc\'\i a, Granada, Spain}
\altaffiltext{3}{Department of Astronomy, Univ. of Alabama, Tuscaloosa, USA}


\begin{abstract}
Although it is widely accepted that galaxy interactions stimulate secular evolutionary effects 
(e.g. enhanced star formation) the amplitude of this effect and the 
processes for accomplishing them, 
are not well quantified.  The goal of the project AMIGA (Analysis of the Interstellar Medium of 
Isolated Galaxies) is to provide a sizable reference sample (n=1050)  of the most isolated galaxies 
as a basis for the study of the influence of the environment on galaxy properties.
Here, we present the far-infrared (FIR) properties of 1030 galaxies of the sample for which IRAS 
data are available. We improved the detection rate and accuracy of the IRAS data with respect to 
the Point Source and Faint Source Catalog by redoing the data reduction with the IPAC utility  
ADDSCAN/SCANPI. Comparing the FIR to the blue luminosities, we find a slightly non-linear
relation. Furthermore, we find  that interacting galaxies tend to
have an enhanced FIR emission. 
\end{abstract}



\section{The AMIGA project}

A key question in astrophysics is the relative role of nurture versus nature in 
galaxy evolution. In order to make progress, studies need to be based on a well-defined sample
of isolated galaxy which has been lacking so far.   
We are compiling and analysing data for the first complete unbiased 
control sample of the most isolated galaxies in the northern sky (Leon \& Verdes-Montenegro 2003, 
Verdes-Montenegro et al. 2005).
To compare and quantify the properties of  different phases of the 
interstellar medium, as well as the  level of star formation, we are building 
a multiwavelength database (far-infrared, near-infrared, 
optical, H$\alpha$, radio continuum, HI and CO) for this sample.
The data will be  publicly available from www.iaa.es/AMIGA.html. 

Our sample is based on the Catalogue of Isolated Galaxies (CIG, 1050 galaxies, 
Karatchenseva 1973 ) assembled with the requirement that no similarly sized galaxies with 
diameter d (where d is between 1/4 and 4 times diameter D of the CIG galaxy) lie within 20d. 
We chose the CIG as a basis 
because this sample  presents various advantages: (i) It is selected using a powerful criterium,
so that 
the CIG contains a large fraction of the most isolated nearby galaxies 
in the northern hemisphere.
Since the  selection criterium does not take into account redshift, it actually
excludes some galaxies which have only apparent companions that lie in reality at a very different
redshift. This is however not a problem for our purpose because 
(ii) the sample is large enough to be statistically significant. It furthermore
covers a large enough volume to be almost (80\%) optically complete
up to a Zwicky magnitude of 15 mag (Verdes-Montenegro et al. 2005).
(iii) Finally, the fact that the galaxies in the CIG are nearby (the bulk of the galaxies have
recession velocities below 10\,000 km/s) enables us to determine the morphologies
in a reliable way (Sulentic et al. 2006).  Since furthermore 
all morphological types are found in CIG, 
we are able to study galaxy properties as a function of galaxy type.

As a first step, we are performing a number of refinements to the CIG:
a) We are carrying out a computational 
revision and quantification  of the degree of
isolation using SExtractor and LMORFO to the POSSI plates (Verley et al. in prep.), 
b) we are revising the 
morphologies with the help of POSSII and our optical images 
(Sulentic et al. 2006), and c) we have checked the positions and accumulated new redshifts available 
in the literature (Leon \& Verdes-Montenegro 2003, Verdes-Montenegro et al. 2005).

\section{Reprocesssing of IRAS data}

We obtained the IRAS fluxes at 12, 25, 60 and 100 $\mu$m using the ADDSCAN/SCANPI utility at IPAC.  
We followed the recommendation for the calculation of the total fluxes and visually inspected 
all spectra in order to check for  (i) the presence of cirrus emission, (ii) confusing with 
neighboring galaxies and (iii) the significance of the detection (e.g. confusing with noise spikes).
A more detailed description  of the data processing will be presented in Lisenfeld 
et al. (in preperation).
This reprocessing yielded:

\vspace{-0.1cm}
\begin{itemize}
\item An increase in the number of data points in comparison to the IRAS  Point 
Source Catalog (PSC) 
and Faint Source Catalog (FSC): Whereas there are only 524 galaxies of the 1050 CIG galaxies in 
the PSC/FSC, the ADDSCAN/SCANPI reduction provided data for 1031 objects.

\vspace{-0.1cm}
\item An improvement of the signal-to-noise-ratio by a factor of 2-5. 
In particular, (55, 70, 9, 81) galaxies at (12, 25, 60, 100)$\mu$m
were only upper limits in the PSC/FSC but changed to detections after our
reprocessing.

\vspace{-0.1cm}
\item An improved accuracy of the fluxes, because ADDSCAN/SCANPI 
is able to measure the total flux of 
extended objects, as long as their size is not above a few arcmin. 
We found a 
trend of the ratio of the flux derived with  ADDSCAN/SCANPI to the flux
from the PSC/FSC to increase with source diameter, especially at short wavelengths,
 suggesting that the fluxes
in the PSC/FSC indeed underestimate the correct fluxes for large object.
Furtermore, our visual
inspection of the spectra allowed us to reject dubious cases.
In fact, we classified (29, 21, 5, 3) galaxies at  (12, 25, 60, 100)$\mu$m as
non-detection that were listed as detection in the PSC/FSC.
\end{itemize}

\section{Relation between L$_{\rm FIR}$ and  L$_{\rm B}$}

As a first result, we show in Fig. 1 a comparison of 
the FIR luminosity, L$_{\rm FIR}$,  
(calculated from the 60 and 100 µm fluxes) to the 
blue luminosity, L$_{\rm B}$,  derived from the corrected Zwicky magnitudes 
(see Verdes-Montenegro et al. 2005). 
A more detailed analysis of the data, including the full
presentation  of the
characteristics of the FIR luminosities and colours will be presented in Lisenfeld et al. 
(in prepartion).
We limit the sample to 736 galaxies with
optical magnitudes between 11 and 15 mag, representing an 80\%  
complete subsample of the 
CIG  (Verdes-Montenegro et al. 2005). Furthermore, based on the morphological revision 
of the sample we exclude 23 galaxies which are judged to be interacting  (Sulentic et al. 
2006).

\begin{figure}
\centerline{
\includegraphics[width=6cm]{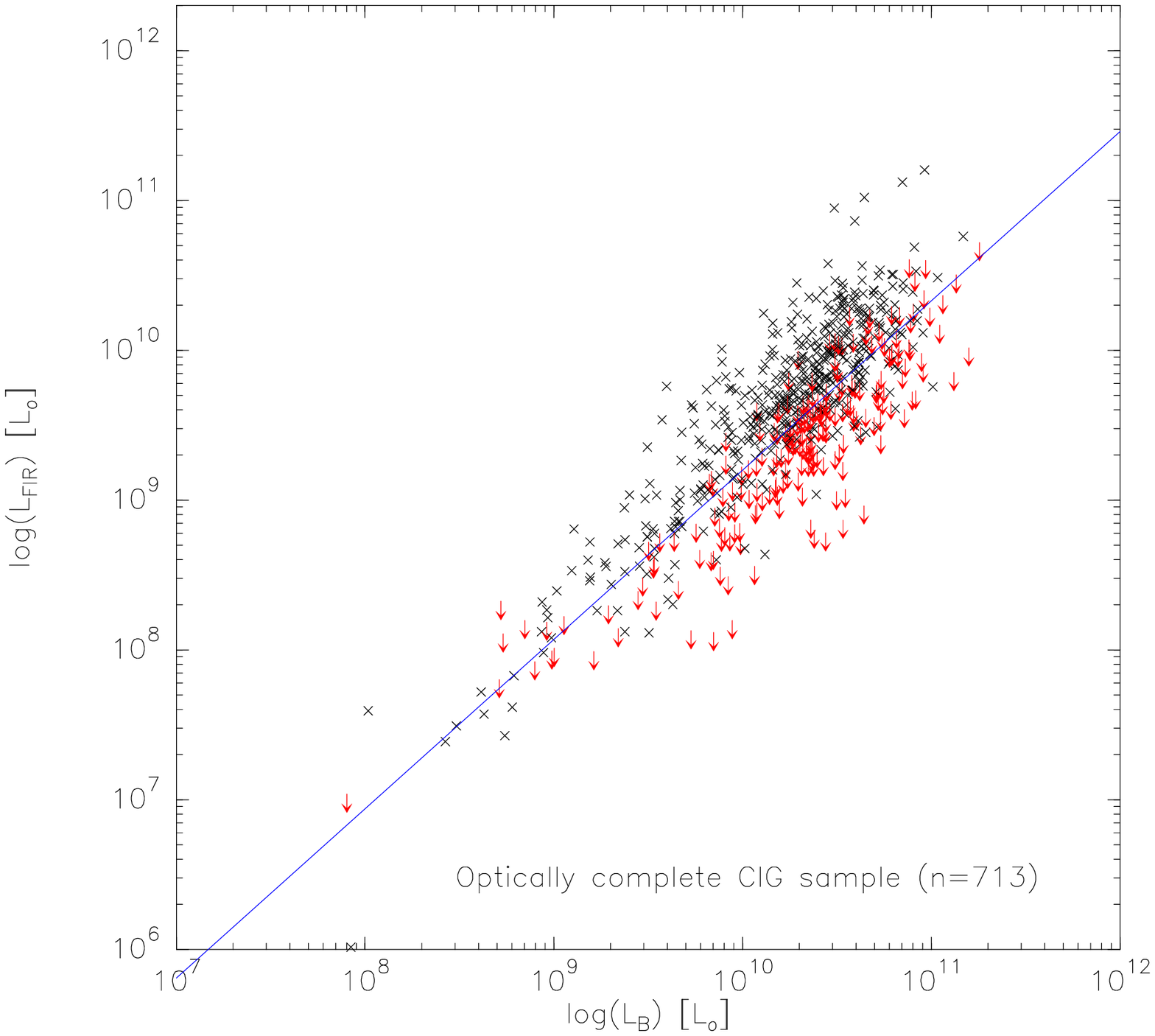} 
\includegraphics[width=6cm]{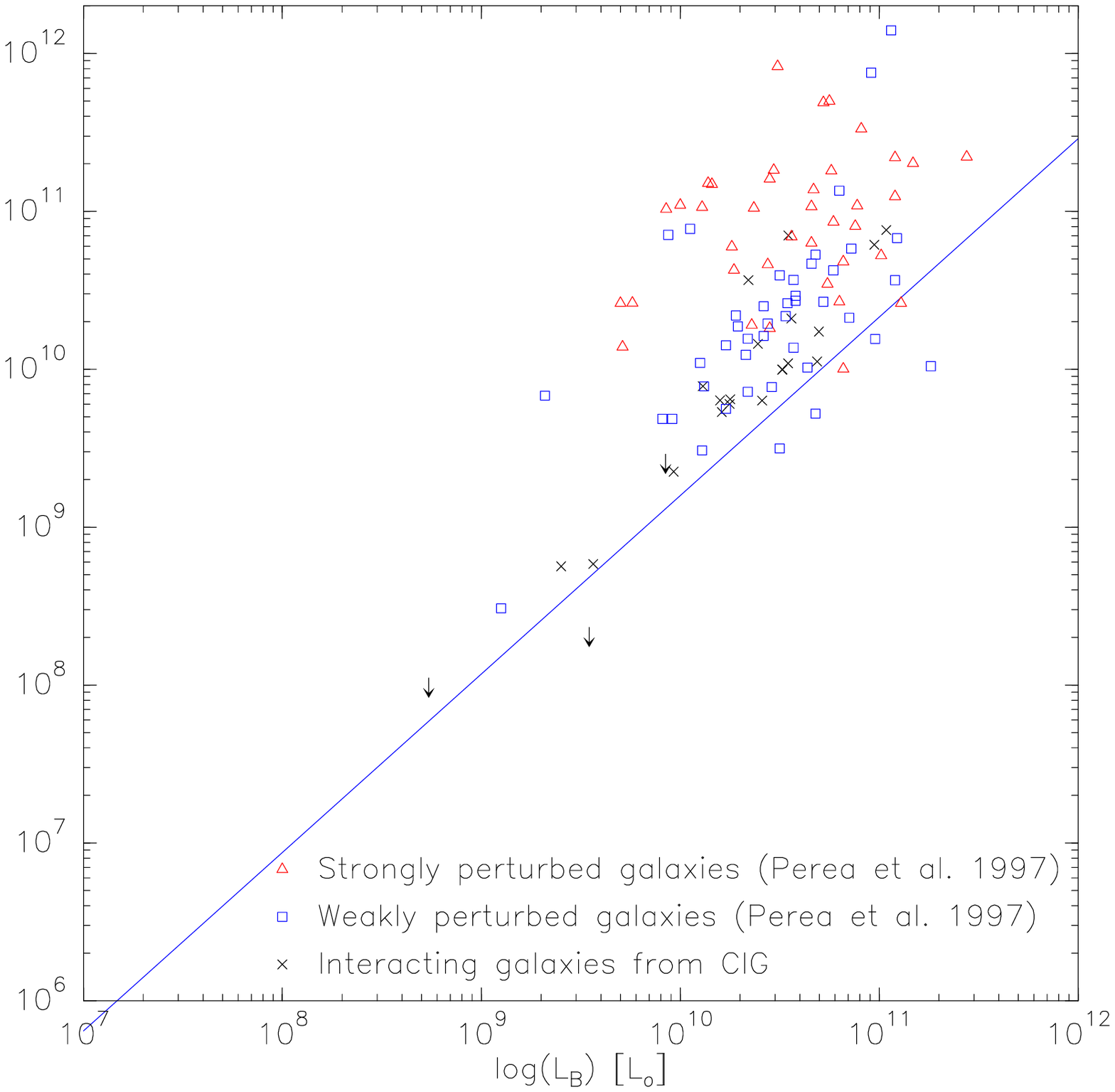} 
}
\caption{
The relation between the FIR and blue luminosity for
an optically complete 
subsample of the
CIG, excluding 23 interacting CIG galaxies {\bf (left)} and 
for different samples of interacting galaxies
{\bf (right)} The line is in both panels the regression found for the 
CIG (eq. 1).
}
\end{figure}

We fit the correlation, taking into account the upper 
limits by applying survival methods from the package ASURV (Feigelson \& Nelson 1985,
Isolbe, Feigelson \& Nelson 1986)
%
 and obtain for the 
relation (adopting  L$_{\rm B}$ as the independent variable). 
\begin{equation}
\log(L_{\rm FIR}) = (1.13 \pm 0.03) \log(L_{\rm B}) - (2.1 \pm 0.3)
\end{equation}
The slope obtained for the sample of the 23 clearly interacting CIGs was considerably higher,
1.46 $\pm$   0.14.  The difference is due to an increase in L$_{\rm FIR}$: Whereas the
average L$_{\rm B}$ of both samples are basically the same ($\nobreak{<L_{\rm B}>} = 
10.22 \pm 0.02$ for 
the 713 CIG galaxies and $\nobreak{< L_{\rm B}>} = 10.23 \pm 0.11$ for 
the 23 interacting CIG galaxies) the FIR luminosity is increased for the interacting galaxies
($\nobreak{< L_{\rm FIR}>} = 9.18 \pm 0.08$ for 
the 713 CIG galaxies and $\nobreak{< L_{\rm B}>} = 9.75 \pm 0.17$ for 
the 23 interacting CIG galaxies).

In Fig. 1 (left) we show different samples of interacting galaxies compared to the
slope of the CIG (eq. 1).
The interacting galaxies
clearly lie above this slope, indicating an enhancement
of the FIR emission compared to L$_{\rm B}$.

\begin{acknowledgments}
UL, LVM and SL are partially supported by DGI (Spain) AYA 2002-03338, AYA2004-08251-CO2-02 (UL),
the Junta de Andaluc\'\i a and the Universidad de  Granada.
\end{acknowledgments}



%




\end{document}